\documentclass[11pt]{article}
\usepackage{cite}
\usepackage{mdframed}
\usepackage{epsfig,subfigure}
\usepackage{amssymb}
\usepackage[fleqn]{amsmath}
\usepackage{color}
\usepackage{arydshln}
\usepackage{graphicx}
\usepackage{color}
\usepackage[dvipsnames]{xcolor}

\definecolor{armygreen}{rgb}{0.29, 0.33, 0.13}

\usepackage{amssymb}
\usepackage{amsmath,amsfonts,amssymb}
\usepackage{verbatim}
\usepackage{stfloats}
\usepackage[bookmarks=false]{}
\usepackage{algorithm}
\usepackage{algcompatible}

\newtheorem{theorem}{Theorem}

\newtheorem{definition}{Definition}
\newtheorem{lemma}{Lemma}

\newtheorem{remark}{Remark}

\newcommand\blfootnote[1]{%
  \begingroup
  \renewcommand\thefootnote{}\footnote{#1}%
  \addtocounter{footnote}{-1}%
  \endgroup
}

\begin{document}
\date{}

\title{
On the Linear Capacity of Conditional Disclosure of Secrets
}
\author{\normalsize Zhou Li and Hua Sun \\
}

\maketitle

\blfootnote{
Zhou Li (email: zhouli@my.unt.edu) and Hua Sun (email: hua.sun@unt.edu) are with the Department of Electrical Engineering at the University of North Texas. }

\maketitle

\begin{abstract}
Conditional disclosure of secrets (CDS) is the problem of disclosing as efficiently as possible, one secret from Alice and Bob to Carol if and only if the inputs at Alice and Bob satisfy some function $f$.
The information theoretic capacity of CDS is the maximum number of bits of the secret that can be securely disclosed per bit of total communication. 
All CDS instances, where the capacity is the highest and is equal to $1/2$, are recently characterized through a noise and signal alignment approach and are described using a graph representation of the function $f$. In this work, we go beyond the best case scenarios and further develop the alignment approach
to characterize the linear capacity of a class of CDS instances to be $(\rho-1)/(2\rho)$, where $\rho$ is a covering parameter of the graph representation of $f$. 
\end{abstract}

\newpage

\allowdisplaybreaks
\section{Introduction}
\begin{figure}[h]
\begin{center}
\includegraphics[width= 4 in]{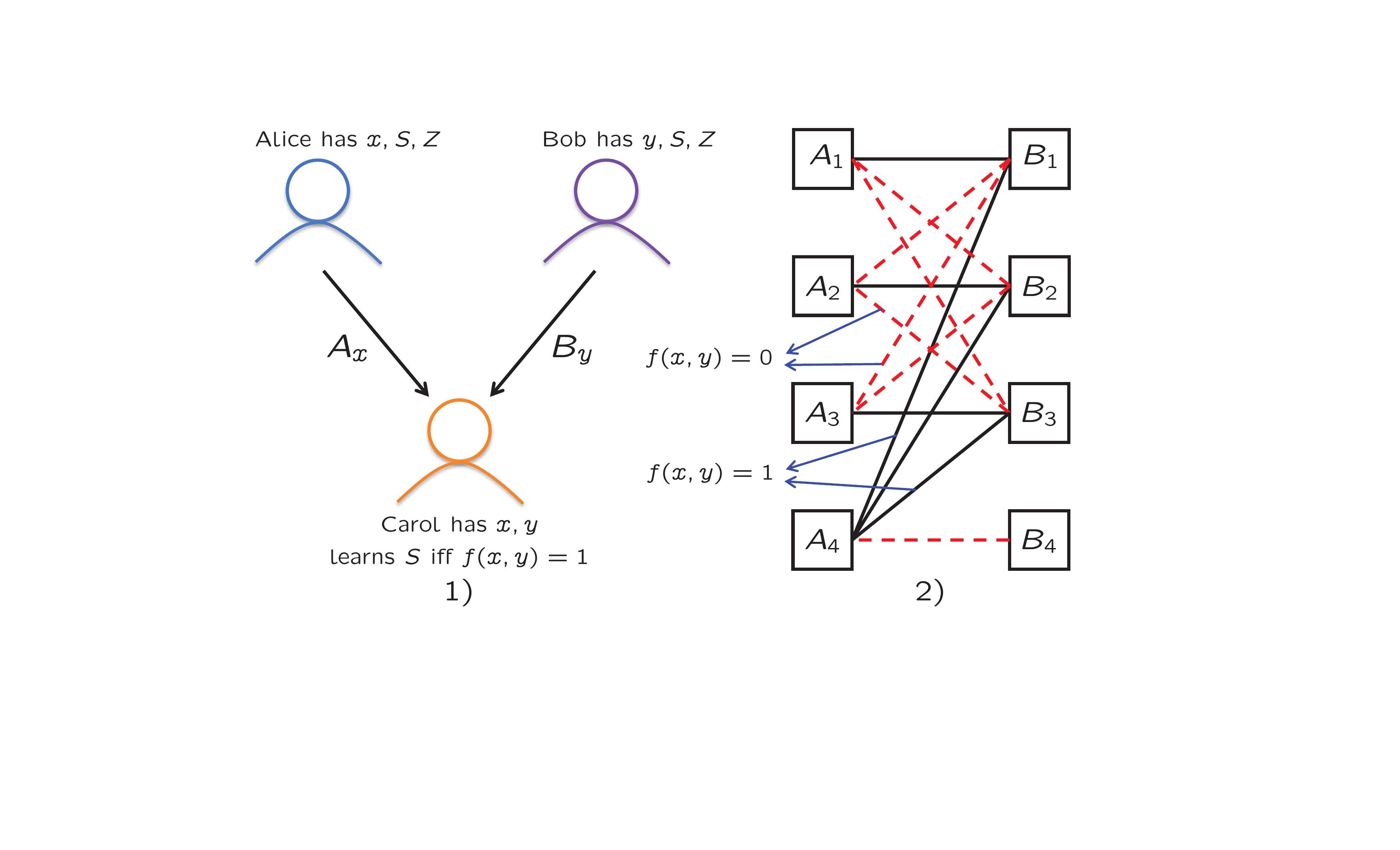}
\caption{\small 1). Alice and Bob (with secret $S$, noise variable $Z$, respective inputs $x,y$) wish to disclose the secret $S$ to Carol if and only if $f(x,y) = 1$ for a binary function $f$, through signals $A_x, B_y$. 2) An example of $f(x,y)$ in graph representation. 
From pair of nodes connected by a solid black edge (i.e., $f(x,y) = 1$), Carol can decode $S$; from pair of nodes connected by a dashed red edge (i.e., $f(x,y) = 0$), Carol learns nothing about $S$ in the information theoretic sense.
}
\label{fig:prob}
\end{center}
\end{figure}

The conditional disclosure of secrets (CDS) problem is a classical cryptographic primitive with rich connections to many other primitives such as symmetric private information retrieval \cite{SymPIR} and secret sharing \cite{liu2018towards, applebaum2020better}. For more background and applications of CDS, we refer to the introduction section of \cite{Li_Sun_CDS} and references therein. The goal of the CDS problem is to find the most efficient way for Alice and Bob to disclose a common secret to Carol if and only if the inputs at Alice and Bob satisfy some function $f$ (see Fig.~\ref{fig:prob}). The CDS problem was initially studied in the setting where the secret is one bit long, and the cost of a CDS scheme is measured by the worst case total amount of communication over all functions $f$, typically as order functions of the input size \cite{SymPIR, Gay_Kerenidis_Wee, Applebaum_Arkis_Raykov_Vasudevan, Laur_Lipmaa, Applebaum_Vasudevan, Liu_Vaikuntanathan_Wee}. That is, the focus is on the scaling law of the communication complexity as the input size grows to infinity.
What is pursued in this work is the traditional Shannon theoretic formulation, where the secret size is allowed to be arbitrarily large, and the communication rate is the number of bits of the secret that can be securely disclosed per bit of total communication. The aim is to characterize the maximum rate, termed the capacity of CDS, for a fixed function $f$.

\vspace{0.2in}

In \cite{Li_Sun_CDS}, we obtain a complete characterization for all functions $f$ where the CDS capacity is the highest, and is equal to $1/2$. In describing this result, we find it convenient to represent the function $f$ by a bipartite graph, where each node denotes a possible signal for certain input and two types (colors) of edges are used to denote whether $f$ is $1$ or $0$ (see Fig.~\ref{fig:prob}.2). We will use this graph representation of functions $f$ throughout this work. The feasibility condition for capacity $1/2$ is then stated in terms of the graphic properties of $f$. Furthermore, this result is obtained using a novel noise and signal alignment approach, which guides the proof of both (information theoretic) impossibility claims and (linear) protocol designs.

Beyond the best rate scenarios, the simplest uncovered case is also considered in \cite{Li_Sun_CDS} (see Theorem~2), where the linear capacity\footnote{It turns out that the linear capacity, i.e., the highest rate achievable by linear schemes, does not match the best converse bound produced by only Shannon information inequalities, i.e., sub-modularity of entropy functions \cite{Li_Sun_CDS}.} has been found and this is our starting point. Our goal in this work is to further develop the alignment approach to characterize the linear capacity of a larger class of CDS instances. As our first main result (see Theorem~\ref{thm:con}), we obtain a general converse bound for linear CDS schemes, which applies to any CDS instance, is parameterized by a covering parameter $\rho$ of the graph representation of $f$, and is equal to $(\rho-1)/(2\rho)$. As our second main result (see Theorem~\ref{thm:ach}), we show that the above converse bound is achievable for a class of graphs, i.e., CDS instances, through a vector linear code based achievable scheme with matching rate. While we find that the converse bound appears to be achievable for more graphs (by verifying a number of examples), an explicit condition of a larger class and a universal code design that applies generally remain elusive. As our final result, we show through an example that the above converse bound is not tight in general and we establish the linear capacity for that example (see Theorem~\ref{thm:ex}). Interestingly, all results are obtained through a more refined view of the alignment approach.


\section{Problem Statement and Preliminaries}\label{sec:model}
Consider a binary function $f(x,y)$, where $(x,y)$ is from some set $\mathcal{I} \subset \{1,2,\cdots, X\} \times \{1,2,\cdots, Y\}$ and its characteristic undirected bipartite graph $G_f = (V, E)$, where the node set $V = \{A_1, \cdots, A_X$, $B_1, \cdots, B_Y\}$ and the edge set $E$  is comprised of the unordered pairs $\{A_x, B_y\}$ such that $(x, y) \in \mathcal{I}$. The edges have two types: if $f(x,y) = 1$, $\{A_x, B_y\}$ is a solid black edge and is referred to as a {\em qualified edge};  if $f(x,y) = 0$, $\{A_x, B_y\}$ is a dashed red edge and is referred to as an {\em unqualified edge} (see Fig.~\ref{fig:prob}.2 for an example).

The variable $x$ ($y$) denotes the input available only to Alice (Bob) and $A_x$ ($B_y$) denotes the signal sent from Alice (Bob) to Carol for securely disclosing the secret $S$, which is comprised of $L$ i.i.d. uniform symbols from a finite field $\mathbb{F}_p$. In addition to the secret $S$, Alice and Bob also hold an independent common noise variable $Z$ (to assist with the secure disclosure task) that is comprised of $L_Z$ i.i.d. uniform symbols from $\mathbb{F}_p$. In $p$-ary units,
\begin{eqnarray}
H(S) = L, ~H(Z) = L_Z, ~H(S, Z) = H(S) + H(Z) = L + L_Z. \label{sz_ind}
\end{eqnarray}

Each signal $A_x$ ($B_y$) is assumed to be comprised of $N$ symbols from $\mathbb{F}_p$ and must be determined by information available to Alice (Bob).
\begin{eqnarray}
H(A_x, B_y | S, Z) = 0. \label{det}
\end{eqnarray}

The disclosure task is said to be successful if the following conditions are satisfied. From a qualified edge, Carol can recover $S$ with no error; 
from an unqualified edge, Carol must learn nothing about $S$. For all $(x,y) \in \mathcal{I}$, we have
\begin{eqnarray}
&& [\mbox{Correctness}] ~~H(S | A_x, B_y) = 0,  ~~~~~~~~~\mbox{if}~f(x,y) = 1; \label{dec} \\
&& [\mbox{Security}] ~~~~~~H(S | A_x, B_y) = H(S), ~~~\mbox{otherwise}~f(x,y) = 0. \label{sec}
\end{eqnarray}
The collection of the mappings from $x,y,S,Z$ to $A_x, B_y$ as specified above is called a CDS scheme.


The CDS rate $R$ characterizes how many symbols of the secret are securely disclosed per symbol of total communication and is defined 
as follows. 
\begin{eqnarray}
R = \frac{L}{2N}. 
\label{rate}
\end{eqnarray}
A rate $R$ is said to be achievable if there exists a CDS scheme, for which the correctness and security constraints (\ref{dec}), (\ref{sec}) are satisfied and the rate is greater than or equal to $R$. The supremum of achievable rates is called the capacity of CDS, $C$. 


In this work, we focus mainly on the metric of capacity $C$ and allow as much noise as needed, i.e., the randomness size $L_Z$ is unconstrained.

\subsection{Graph Definitions}
We will use some graphic notions of $G_f = (V, E)$ to state our results, which are defined as follows. 
Without loss of generality, we assume that for any node $v \in V$, there exists some node $u \in V$ such that $\{u,v\} \in E$ is an unqualified edge (otherwise, for any $v$ that is connected to only qualified edges, we can set $v$ to be the secret $S$ and then eliminate $v$ and its edges). 

\begin{definition}[Qualified/Unqualified Path/Component]
A sequence of distinct connecting qualified (unqualified) edges is called a {\em qualified (unqualified) path}. A {\em qualified (unqualified) connected component} is a maximal induced subgraph of $G_f$ such that any two nodes in the subgraph are connected by a qualified (unqualified) path.
\end{definition}

For example, in Fig.~\ref{fig:prob}.2, $P = \left\{\{A_1, B_2\}, \{B_2, A_3\}, \{A_3, B_1\}\right\}$ is an unqualified path; as the graph is connected, it is a qualified component.

\begin{definition}[Internal Qualified Edge]
A qualified edge that connects two nodes in an unqualified path is called an {\em internal qualified edge}.
\end{definition}

For example, in Fig.~\ref{fig:prob}.2, the edge $e = \{A_1, B_1\}$ is an internal qualified edge that connects the two nodes $A_1, B_1$ in the unqualified path $P = \left\{\{A_1, B_2\}, \{B_2, A_3\}, \{A_3, B_1\}\right\}$.

\begin{definition}[Connected Edge Cover]\label{def:ec}
Consider an internal qualified edge $e$ in an unqualified path $P$ and the node set of $P$ is denoted as $V_P \subset V$. A {\em connected edge cover} of $V_P$ is a set of connected\footnote{That is, any two nodes in $M$ are connected by a qualified path.} qualified edges $M \subset E$ such that each node in $V_P$ is covered by at least one qualified edge in $M$ and $e\in M$. The size of a connected edge cover for $(e,P)$ is the number of edges in $M$ and is denoted as $\rho(e, P)$. If no such $M$ exists, then $\rho(e, P)$ is defined as $+\infty$. Further, $\rho \triangleq \min_{e, P} \rho(e, P)$.
\end{definition}

For example, in Fig.~\ref{fig:prob}.2, consider the internal qualified edge $e = \{A_1, B_1\}$ in the unqualified path $P = \left\{\{A_1, B_2\}, \{B_2, A_3\}, \{A_3, B_1\}\right\}$, then the nodes in $P$ are $V_P = \{A_1, B_2, A_3, B_1\}$ and a connected edge cover of $V_P$ is $M = \{\{A_4, B_1\}, \{A_4, B_2\}, \{A_4, B_3\}, \{A_1, B_1\}, \{A_3, B_3\}\}$. In this case, $\rho(e, P) = 5$ as $M$ contains $5$ edges and we can verify that the minimum value of $\rho(e, P)$ over all internal qualified edges and their associated unqualified path pairs $(e, P)$ is $\rho = 5$. 

It can be verified that in general, $\rho$ can be any integer that is at least $5$. Also note that as $\rho$ is defined to be the minimum over all $e, P$, so the connected edge cover $M$ that attains the value of $\rho$ corresponds to one that has the minimal cardinality.
 
\subsection{Linear Feasibility}\label{sec:linear}
We characterize the feasibility condition of a linear CDS scheme.

\bigskip
\noindent {\bf Linear Scheme:} {\it For a feasible linear CDS scheme, each signal (equivalently, each node $v \in V$) 
\begin{eqnarray}
v = {\bf F}_v S + {\bf H}_v Z, ~{\bf F}_v \in \mathbb{F}_p^{N \times L}, {\bf H}_v \in \mathbb{F}_p^{N \times L_Z} \label{eq:linear}
\end{eqnarray}
is specified by two 
precoding matrices, ${\bf F}_v$ for the secret $S \in \mathbb{F}_p^{L\times 1}$ and ${\bf H}_v$ for\footnote{Without loss of generality, we assume that ${\bf H}_v$ has full row rank, i.e., $\mbox{rank}({\bf H}_v) = N$, because each $v$ is assumed to connect to at least an unqualified edge so that $I(v; S) = 0$, then the linearly dependent rows of ${\bf H}_v$ in $v$ must be linearly dependent as well (thus redundant).} the noise $Z \in \mathbb{F}_p^{L_Z \times 1}$ such that the following properties are satisfied.
\begin{itemize}
\item Consider any edge $\{v,u\}$ and identify the overlap of their noise spaces, i.e., the row space of ${\bf H}_v$ and ${\bf H}_u$. That is, find matrices ${\bf P}_v$ and ${\bf P}_u$ 
such that
\begin{eqnarray}
{\bf P}_v {\bf H}_v &=& {\bf P}_u {\bf H}_u, \notag\\
\mbox{rank}({\bf P}_v) &=& \mbox{rank}({\bf P}_u) = \mbox{dim}(\mbox{rowspan}({\bf H}_v) \cap \mbox{rowspan}({\bf H}_u)), \label{eq:proj}
\end{eqnarray}
then the secret spaces satisfy
\begin{eqnarray}
&& [\mbox{Correctness}] ~~\mbox{rank}({\bf P}_v {\bf F}_v - {\bf P}_u {\bf F}_u) = L,  ~~~\mbox{if $\{u,v\}$ is qualified}; \label{eq:dec} \\
&& [\mbox{Security}]\footnotemark
~~~~~~{\bf P}_v {\bf F}_v = {\bf P}_u {\bf F}_u, ~~~~~~~~~~~~~~~\mbox{otherwise $\{u,v\}$ is unqualified}. \label{eq:sec}
\end{eqnarray}
\end{itemize}
}

\footnotetext{As a straightforward corollary, note that the security constraint applies to subspaces of the overlapping noise spaces. That is, when $\{u,v\}$ is unqualified,  for all matrices ${\bf \bar{P}}_v, {\bf \bar{P}}_u$ where ${\bf \bar{P}}_v {\bf H}_v = {\bf \bar{P}}_u {\bf H}_u$, we have ${\bf \bar{P}}_v {\bf F}_v = {\bf \bar{P}}_u {\bf F}_u$.} 

The correctness constraint (\ref{eq:dec}) and the security constraint (\ref{eq:sec}) for linear schemes imply the entropic versions (\ref{dec}), (\ref{sec}). For correctness, note that ${\bf P}_v v - {\bf P}_u u = ({\bf P}_v {\bf F}_v - {\bf P}_u {\bf F}_u) S$, so $S$ can be decoded with no error if ${\bf P}_v {\bf F}_v - {\bf P}_u {\bf F}_u$ has full rank. 
For security, note that ${\bf P}_v {\bf H}_v, {\bf P}_u {\bf H}_u$ contains all the overlaps so that the remaining vectors are orthogonal. That is,
\begin{eqnarray}
({\bf H}_v; {\bf H}_u) \overset{\mbox{\scriptsize invertible}}{\longleftrightarrow} ({\bf P}_v {\bf H}_v; {\bf Q}_v {\bf H}_v; {\bf Q}_u {\bf H}_u) \label{eq:inv}
\end{eqnarray}
where $\mbox{rowspan}({\bf P}_v {\bf H}_v),~\mbox{rowspan}({\bf Q}_v {\bf H}_v),~\mbox{rowspan}({\bf Q}_u {\bf H}_u)$ are linearly independent. Then we have
\begin{eqnarray}
I(S ; v, u) &\overset{(\ref{eq:proj}) (\ref{eq:sec})}{=}& I(S ; {\bf P}_v {\bf F}_v S + {\bf P}_v {\bf H}_v Z, {\bf Q}_v {\bf F}_v S + {\bf Q}_v {\bf H}_v Z, {\bf Q}_u {\bf F}_u S + {\bf Q}_u {\bf H}_u Z) \label{eq:e1}\\
&\overset{(\ref{sz_ind})}{=}& H({\bf P}_v {\bf F}_v S + {\bf P}_v {\bf H}_v Z, {\bf Q}_v {\bf F}_v S + {\bf Q}_v {\bf H}_v Z, {\bf Q}_u {\bf F}_u S + {\bf Q}_u {\bf H}_u Z) \notag\\
&&~- H( {\bf P}_v {\bf H}_v Z, {\bf Q}_v {\bf H}_v Z, {\bf Q}_u {\bf H}_u Z ) \\
&\leq& \mbox{rank}({\bf P}_v {\bf H}_v; {\bf Q}_v {\bf H}_v; {\bf Q}_u {\bf H}_u) - \mbox{rank}({\bf P}_v {\bf H}_v; {\bf Q}_v {\bf H}_v; {\bf Q}_u {\bf H}_u) = 0
\end{eqnarray}
where (\ref{eq:e1}) follows from the fact that ${\bf P}_v {\bf F}_v S + {\bf P}_v {\bf H}_v Z = {\bf P}_u {\bf F}_u S + {\bf P}_u {\bf H}_u Z$ (see (\ref{eq:proj}), (\ref{eq:sec})) and linear transformation to identify the overlap is invertible (see (\ref{eq:inv})). In the last step, the first term follows from counting the number of variables and the property that uniform distribution maximizes entropy, and the second term follows from the fact the symbols in $Z$ are i.i.d. and uniform.
Conversely, any feasible linear scheme must satisfy (\ref{eq:dec}), (\ref{eq:sec}).
Such a linear feasibility framework has appeared in related problems, e.g., index coding \cite{Dau_Skachek_Chee_Security} and secure groupcast \cite{Sun_GenercSG}. 

To facilitate later use, we summarize some useful properties of feasible linear schemes in the following lemma. 
A detailed proof can be found in Lemma 6 and Lemma 7 of \cite{Li_Sun_CDS}.

\begin{lemma}\label{lemma:align}
For any linear scheme as defined above and any edge $\{v,u\}$, we have
\begin{eqnarray}
&& [\mbox{Noise Alignment}] ~~
\mbox{dim}(\mbox{rowspan}({\bf H}_v) \cap \mbox{rowspan}({\bf H}_u)) \geq L,  ~~~\mbox{if $\{u,v\}$ is qualified}; \label{eq:noise} \\
&& [\mbox{Signal Alignment}] ~~~~~~~~~~~~~{\bf P}_v {\bf F}_v = {\bf P}_u {\bf F}_u, ~~~~~~~~~~~~~~~~~~~~\mbox{if $\{u,v\}$ is unqualified}. \label{eq:signal}
\end{eqnarray}
\end{lemma}

The intuition of the lemma is as follows. (\ref{eq:noise}) follows from the correctness constraint (\ref{eq:dec}), which requires the overlap of the noise spaces to have at least $L$ dimensions as decoding is only possible over the overlapping space (so referred to as `noise alignment') and other spaces are covered by independent noise variables. (\ref{eq:signal}) follows from the security constraint (\ref{eq:sec}), which says that over the overlapping noise space, the secret space must also be fully overlapping (so referred to as `signal alignment' since both noise and secret fully align in this space) as otherwise the unqualified edge can reveal some linear combination of the secret symbols, violating the security constraint. 

\begin{figure}[h]
\begin{center}
\includegraphics[width= 5 in]{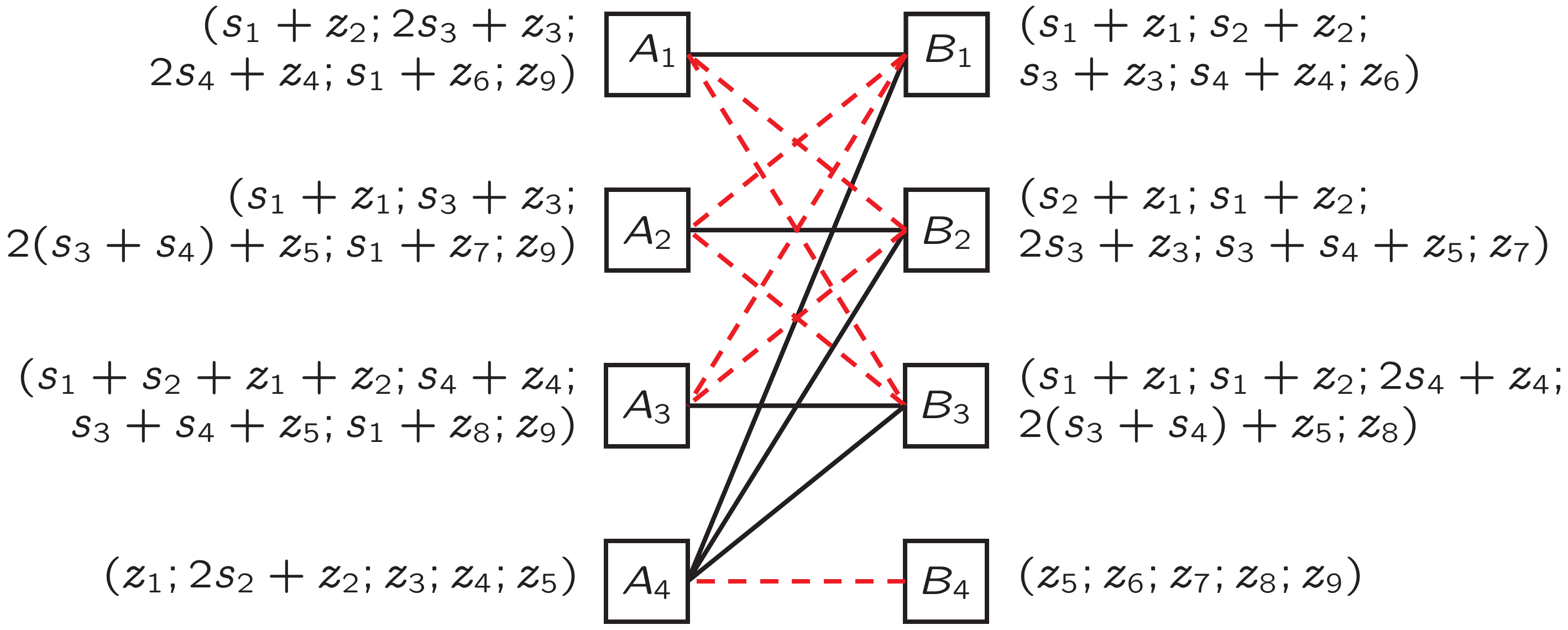}
\caption{\small A CDS instance and the vector linear achievable scheme of rate $R = 2/5$.}
\label{fig:ex1}
\end{center}
\end{figure}

In the remainder of this paper, we use (\ref{eq:dec}) and (\ref{eq:sec}) to verify the correctness and security of a linear scheme. To illustrate how it works, let us consider again the CDS instance in Fig.~\ref{fig:prob}.2 (reproduced in Fig.~\ref{fig:ex1}). We show that rate $R = 2/5$ is achievable, through presenting a vector linear scheme with $L = 4, N = 5$. That is, the secret has $L = 4$ symbols over $\mathbb{F}_3$ ($S = (s_1; s_2; s_3; s_4)$), and each signal has $N=5$ symbols over $\mathbb{F}_3$. The assignment of the signals is given in Fig.~\ref{fig:ex1}. Suppose $Z = (z_1; \cdots; z_9)$, where each $z_i$ is uniform and i.i.d. over $\mathbb{F}_3$.

Let us verify that the above scheme is correct and secure. For simplicity, we do not write out explicitly the precoding matrices ${\bf F}_v$ and ${\bf H}_v$ for a signal $v$ as the dimension is relatively large. Instead, we will directly find the overlap by inspection. Consider qualified edge $\{A_3, B_3\}$. $A_3, B_3$ both contain $(z_1+z_2; z_4; z_5; z_8)$ (noise overlaps) and can then obtain $4$ equations of the secret symbols, $(-s_1+s_2; s_4; s_3+s_4; s_1)$, which can recover $S = (s_1; s_2; s_3; s_4)$. Other cases of qualified edges can be verified similarly. Consider the unqualified edge $\{A_3, B_2\}$. $(z_1+z_2; z_5)$ lies in the overlap of the noise spaces and the secret symbols projecting to this space are both $(s_1 + s_2; s_3 + s_4)$, thus no information is leaked. Other unqualified edges follow similarly. The rate achieved is thus $L/(2N) = 4/10 = 2/5$.

\section{Results}
Our first result is a general converse bound of linear CDS schemes, parameterized by the minimum connected edge cover number of internal qualified edges, $\rho$ and stated in Theorem \ref{thm:con}.

\begin{theorem}\label{thm:con}
For any CDS instance, the following converse bound holds for all linear schemes.
\begin{eqnarray}
R_{\mbox{\scriptsize linear}} \leq \frac{\rho-1}{2\rho}.
\end{eqnarray}
\end{theorem}

\begin{remark}
When $\rho = +\infty$, we have that for any internal qualified edge $e$, there is no set of connected edges that can cover all nodes in the unqualified path containing $e$ (refer to Definition \ref{def:ec}). This is equivalent to that there is no internal qualified edge within any qualified component, which reduces to the feasibility condition of capacity $1/2$ from Theorem 1 of \cite{Li_Sun_CDS}.
\end{remark}

The proof of Theorem \ref{thm:con} is presented in Section \ref{sec:con}.

To illustrate the idea, let us consider the CDS instance in Fig.~\ref{fig:ex1}. Note that $e = \{A_1, B_1\}$ is an internal qualified edge in unqualified path $P = \left\{\{A_1, B_2\}, \{B_2, A_3\}, \{A_3, B_1\}\right\}$, with node set $V_P = \{A_1, B_2, A_3, B_1\}$, where $V_P$ is covered by a connected edge cover $M = \{\{A_4, B_1\}, \{A_4, B_2\}, \{A_4, B_3\}$, $\{A_1, B_1\}, \{A_3, B_3\}\}$ so that $\rho(e, P) = |M| = 5$ and this edge cover number turns out to be the minimum, i.e., $\rho = 5$. Then Theorem \ref{thm:con} indicates that $R_{\mbox{\scriptsize linear}} \leq (\rho-1)/(2\rho) = 2/5$. As rate $2/5$ is linearly achievable (see Fig.~\ref{fig:ex1}), the linear capacity of this CDS instance is $2/5$.
 
The intuition of the linear converse is as follows. (\ref{eq:noise}) in Lemma \ref{lemma:align} gives a lower bound on the dimension of the pairwise noise overlap of the two nodes in a qualified edge. We will start from this pairwise overlap to obtain a lower bound on the dimension of the overlap of all nodes in $M$, i.e., $\mbox{rowspan}({\bf H}_{A_4}) \cap \mbox{rowspan}({\bf H}_{B_1}) \cap \mbox{rowspan}({\bf H}_{B_2}) \cap \mbox{rowspan}({\bf H}_{B_3}) \cap \mbox{rowspan}({\bf H}_{A_1}) \cap \mbox{rowspan}({\bf H}_{A_3}) $, which is the overlap of all pairwise overlaps of the edges in $M$, i.e., $\left(\mbox{rowspan}({\bf H}_{A_4}) \cap \mbox{rowspan}({\bf H}_{B_1})\right) \cap \cdots \cap \left(\mbox{rowspan}({\bf H}_{A_3}) \cap \mbox{rowspan}({\bf H}_{B_3})\right)$. For example, consider the first two edges in $M$.
\begin{eqnarray}
&& \dim\big( \mbox{rowspan}({\bf H}_{A_4}) \cap \mbox{rowspan}({\bf H}_{B_1}) \cap \mbox{rowspan}({\bf H}_{B_2}) \big) \notag\\
&=& \dim\big( \left(\mbox{rowspan}({\bf H}_{A_4}) \cap \mbox{rowspan}({\bf H}_{B_1})\right) \cap \left(\mbox{rowspan}({\bf H}_{A_4}) \cap \mbox{rowspan}({\bf H}_{B_2})\right) \big) \\
&\geq& \dim\big( \mbox{rowspan}({\bf H}_{A_4}) \cap \mbox{rowspan}({\bf H}_{B_1}) \big) + \dim\big( \mbox{rowspan}({\bf H}_{A_4}) \cap \mbox{rowspan}({\bf H}_{B_2}) \big) \notag\\
&&~- \dim\big( \mbox{rowspan}({\bf H}_{A_4}) \big) \label{eq:ee1}\\
&\overset{(\ref{eq:noise})}{\geq}& L + L - N
\end{eqnarray}
where (\ref{eq:ee1}) follows from the sub-modularity of linear rank functions and the direct sum (the space spanned by the union of the two sets of vectors) of the two pairwise overlaps is contained in $\mbox{rowspan}({\bf H}_{A_4})$ as each pairwise overlap involves $\mbox{rowspan}({\bf H}_{A_4})$. The last step follows from the pairwise overlap constraint (\ref{eq:noise}) and the fact that the rank of ${\bf H}_{A_4}$ is $N$. We have now transformed the pairwise overlap of dimension $L$ to $3$-wise overlap of $2L-N$, where a term of $L - N$ is added. 
Next as the edges are connected, we may apply sub-modularity repeatedly and find the overlap of all noise spaces in $M$ by including one connected edge at one time, whose dimension turns out to be no less than $L + (\rho-1) (L-N) = 5L - 4N$, i.e., the $L-N$ term is added $\rho-1$ times (from pairwise overlap to $\rho$-wise overlap). Then by (\ref{eq:signal}) in Lemma \ref{lemma:align}, we know that such noise overlap leads to signal overlap for all nodes $V_P = \{A_1, B_2, A_3, B_1\}$ in the unqualified path $P$, in particular including the two nodes $A_1, B_1$ in the internal qualified edge $e$. As overlapping signal contributes no information  for decoding, such overlap shall not exist (when the noise overlap of $A_1, B_1$ is exactly $L$ in (\ref{eq:noise}) and this will be relaxed in the general proof), i.e., $\rho L - (\rho-1)N \leq 0$, and $R_{\mbox{\scriptsize linear}} = L/(2N) \leq (\rho-1)/(2\rho)$.

\bigskip
Next, we proceed to our second result, which shows that the linear converse in Theorem \ref{thm:con}  is tight for a class of CDS instances and is stated in Theorem \ref{thm:ach}.

\begin{theorem}\label{thm:ach}
For any CDS instance where the qualified edges in each qualified component form either a path or a cycle\footnote{A cycle is a path where the first node is the same as the last node. If the qualified edges form either a path or a cycle, equivalently, we have that each node is connected to at most two qualified edges.}, the linear capacity is $C_{\mbox{\scriptsize linear}} = (\rho-1)/(2\rho)$.
\end{theorem}

Note that Theorem \ref{thm:ach} only places constraints on the structure of qualified edges and works for any possible configuration of unqualified edges.

The proof of Theorem \ref{thm:ach} is presented in Section \ref{sec:ach}. 

\begin{figure}[h]
\begin{center}
\includegraphics[width= 3.3 in]{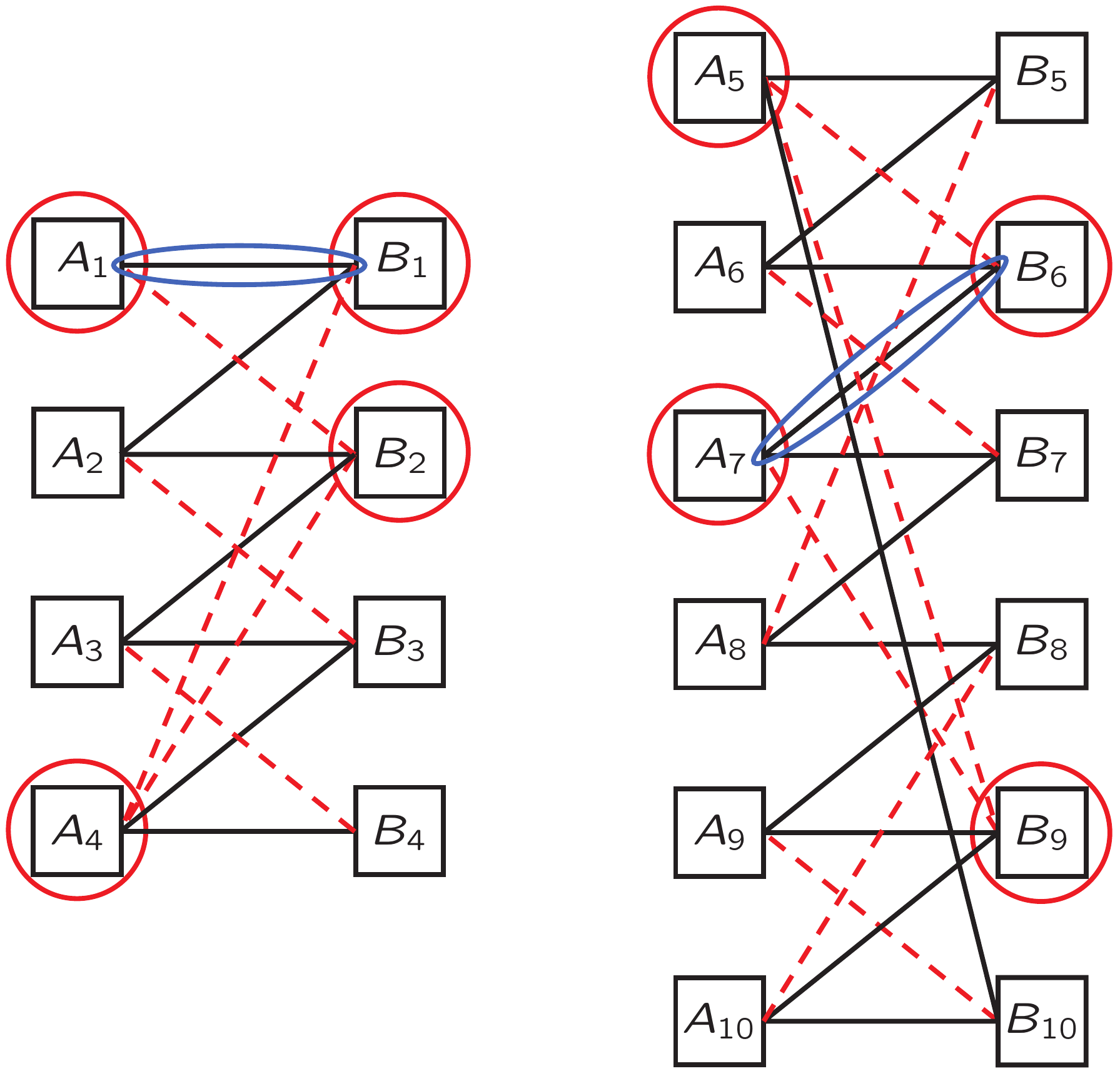}
\caption{\small A CDS instance where each qualified component is either a path or a cycle. For each qualified component, an internal qualified edge $e$ is put in a blue circle and the nodes $V_P$ in the unqualified path $P$ are put in red circles. Then the connected edge cover $M$ is the qualified path that connects to all nodes in $V_P$. For the left qualified component, $\rho(e, P) = 6$ (the path from $A_1$ to $A_4$); for the right qualified component, $\rho(e, P) = 7$ (the path from $B_9$ to $A_7$).
The unqualified edges connecting two nodes from different qualified components are not drawn and can be arbitrary.}
\label{fig:ex2}
\end{center}
\end{figure}

We give an example here to illustrate the idea. Consider the CDS instance in Fig.~\ref{fig:ex2} and we show that the linear capacity is $C_{\mbox{\scriptsize linear}} = 5/12$. Theorem \ref{thm:ach} can be applied as the instance contains two qualified components, where the qualified edges form a path in one qualified component and form a cycle in the other qualified component. $\rho = 6$, because for the left qualified component, there is an internal qualified edge $e = \{A_1, B_1\}$ (see the blue circle) in unqualified path $P = \{\{A_1, B_2\}, \{B_2, A_4\}, \{A_4, B_1\}\}$ (see the red circles), which is then covered by a qualified path with $6$ edges $M = \{ \{A_1,B_1\}, \{B_1, A_2\}, \{A_2, B_2\}, \{B_2, A_3\}, \{A_3, B_3\}, \{B_3, A_4\} \}$. It can be verified that this $M$ has the minimum cardinality, so $\rho = 6$. Then the converse bound follows from Theorem \ref{thm:con}.

\begin{figure}[h]
\begin{center}
\includegraphics[width= 4.5 in]{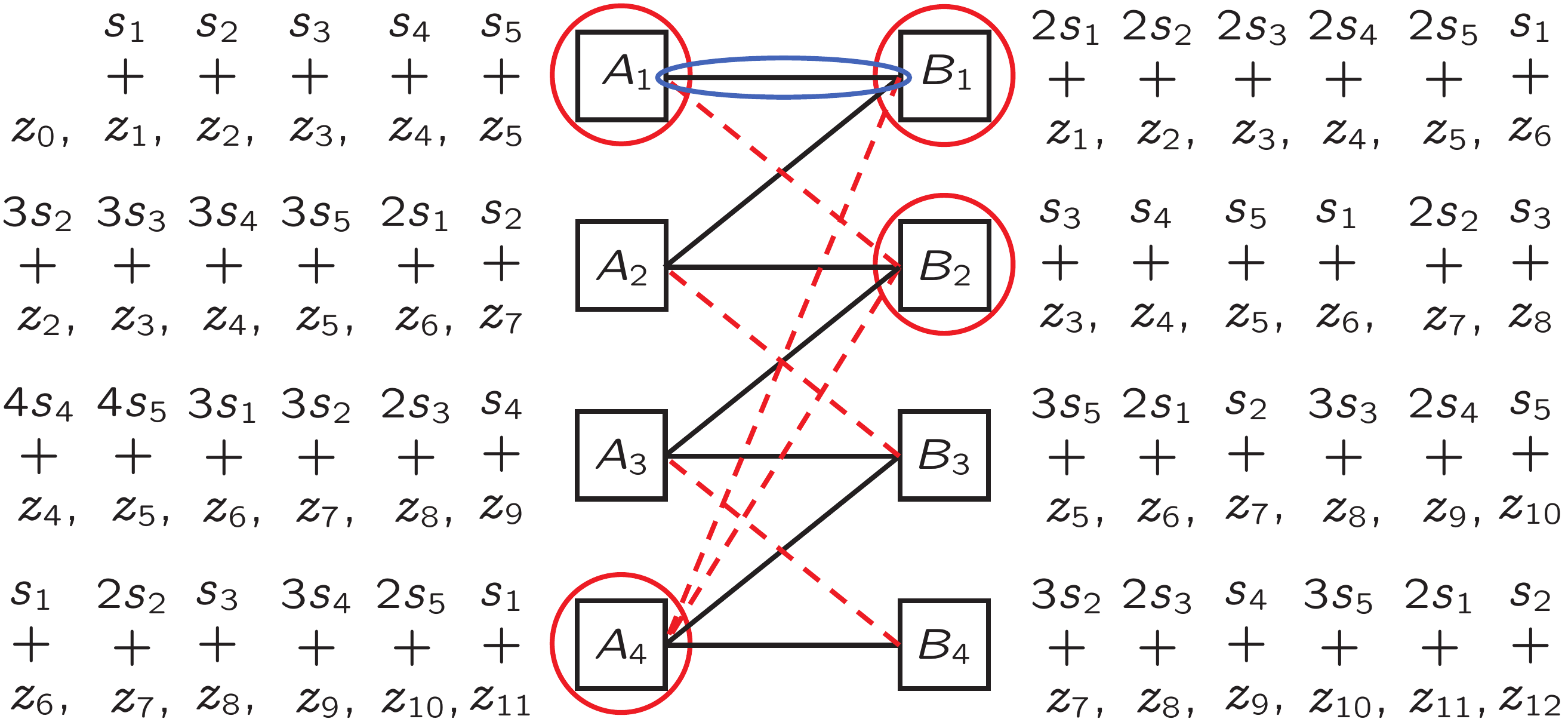}
\caption{\small A linear capacity achieving scheme for the left qualified component.}
\label{fig:ex21}
\end{center}
\end{figure}

We now consider the achievable scheme. We will consider each qualified component one by one and use independent noise variables.
Let us start from the left qualified component (a qualified path from $A_1$ to $B_4$), where the assignment of each signal is given in Fig.~\ref{fig:ex21}. 

The uniform and i.i.d. noise variables are assigned sequentially to the nodes in the path following a sliding window manner, where the first node $A_1$ uses $z_0, z_1, \cdots, z_5$, the second node uses $z_1, \cdots, z_6$, and so on (every two consecutive nodes share $L = 5$ common noise variables). 
Note that this noise assignment does not depend on the structure of the unqualified edges (which is a key property that simplifies the scheme design). The secret symbols $s_1, \cdots, s_5$ are assigned cyclicly to the noise variables, i.e., $(s_1, \cdots, s_5)$ are assigned to $(z_1, \cdots, z_5)$, $(z_6, \cdots, z_{10})$ etc. (i.e., $s_j$ is assigned to $z_{5B + j}$ for any integer $B$). The coefficients of $s_j$ are the only left and most important part. To this end, focus on each $z_i$ in an arbitrary order and consider only the nodes that contain $z_i$. For example, consider $z_6$, which appears in $6$ nodes $B_1, A_2, B_2, A_3, B_3, A_4$ and consider the subgraph induced by these $6$ nodes. For the induced subgraph, consider each unqualified component sequentially (any order will work and one node that connects to no unqualified edge is a trivial unqualified component) and assign the same signal to each node in the unqualified component. So here first consider the unqualified path $\{\{B_1, A_4\}, \{A_4, B_2\}\}$ and assign $s_1 + z_6$ to $B_1, A_4, B_2$; second consider the unqualified path $\{A_2, B_3\}$ and assign $2s_1+z_6$ to $A_2, B_3$; lastly consider $A_3$ and assign $3s_1 + z_6$ to $A_3$. All other $z_i$ can be treated in the same manner (essentially for each $z_i$, we apply the scheme from Theorem 1 of \cite{Li_Sun_CDS}).
This completes the description of the scheme. 

The security and correctness of the scheme follow from the assignment in a straightforward manner. For security, note that for each unqualified edge, the signal for overlapping noise is set to be the same so that the security constraint (\ref{eq:sec}) is satisfied. 
For correctness, we note that in the qualified path, 
every two consecutive nodes are connected by a qualified edge and share $L$ noise symbols. For each shared noise symbol, the secret symbols have different coefficients by noting that the connected edge cover number $\rho > \rho - 1$ and each $z_i$ appears in consecutive $\rho-1$ edges. 

\begin{figure}[h]
\begin{center}
\includegraphics[width= 4.5 in]{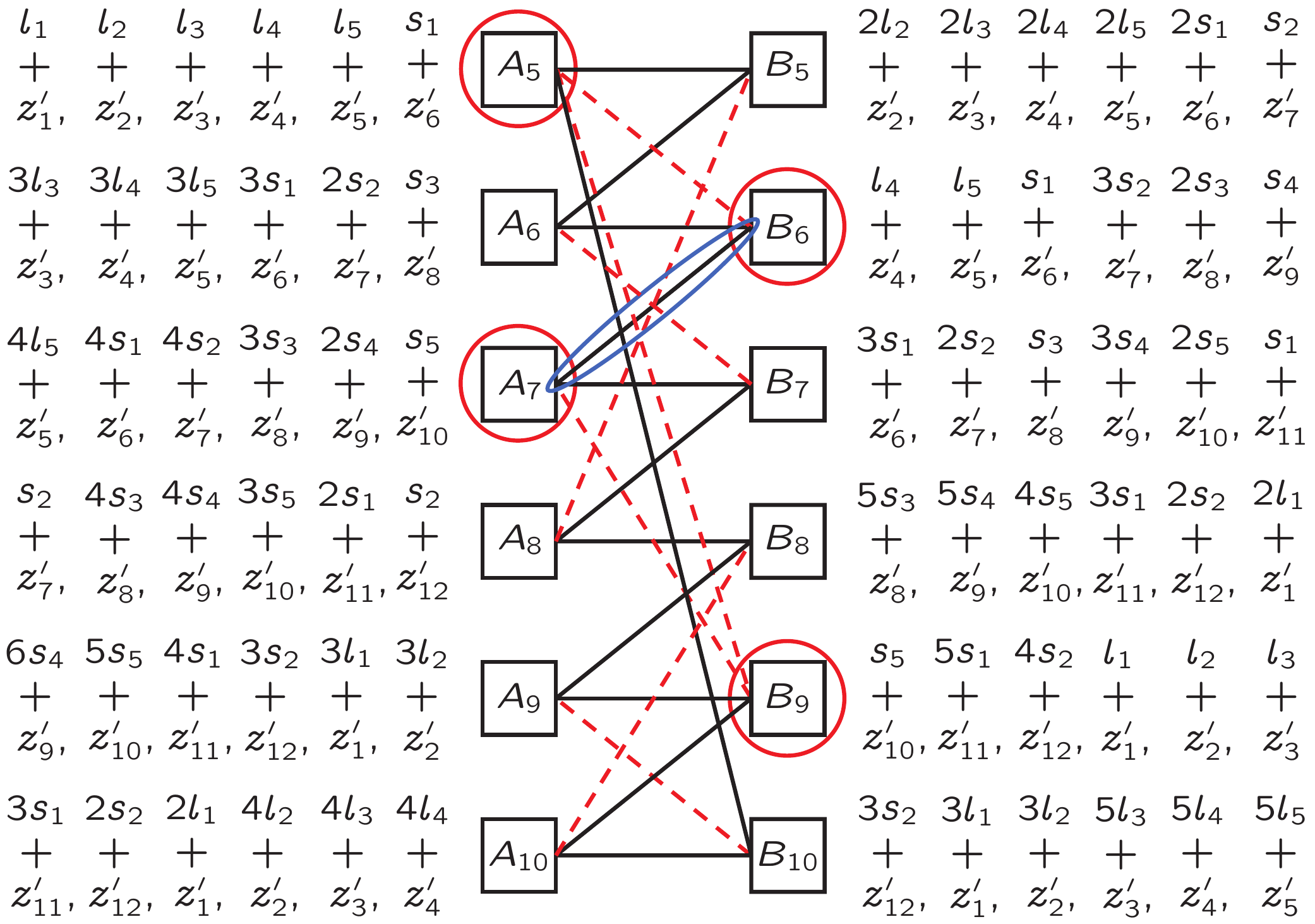}
\caption{\small A linear capacity achieving scheme for the right qualified component.}
\label{fig:ex22}
\end{center}
\end{figure}

After completing the left qualified component, we proceed to the right component (a qualified cycle from $A_5$ to $B_{10}$ and back to $A_5$), whose assignment is given in Fig.~\ref{fig:ex22}. 

The assignment for a cycle is similar to that of a path and we only highlight the differences here. To cope with the fact that the first and last node is the same for a cycle, the sliding window based noise assignment needs to wrap back as well (see the $z_1', \cdots, z_5'$ symbols in Fig.~\ref{fig:ex22}). Also, the secrets associated with nodes near the front and end need to be coded and generic (one linear combination of secret symbols $s_1, \cdots, s_5$ is denoted by $l_i$ in Fig.~\ref{fig:ex22}) so that when combined with any blocks of $s_i$, all secret symbols can be decoded as long as the collective number is sufficient. For example, consider the qualified edge $\{A_{10}, B_9\}$, where we need to recover $S = (s_1, \cdots, s_5)$ from $(s_1, s_2, l_1, l_2, l_3)$.
We will use Cauchy matrix to realize $l_i$ over a sufficiently large field. The other elements for the cycle case are the same as those for a path, i.e., consider each $z_i$ sequentially and set each unqualified component to have the same signal that contains $z_i$ in the induced subgraph. Details are deferred to the general proof.

Lastly, we consider the unqualified edges connecting two nodes from different qualified components, which are not drawn in Fig.~\ref{fig:ex2}. The presence of any number of such unqualified edges will not change the result - for the converse, $\rho$ is not influenced; for the achievability, security is preserved as independent noises are used (i.e., $z_i, z_i'$ in Fig.~\ref{fig:ex21} and Fig.~\ref{fig:ex22}, respectively) for different qualified components. 
The rate achieved is $R = L/(2N) =  5/12$ as the secret has $L=5$ symbols and each signal has $N = 6$ symbols.

\begin{figure}[h]
\begin{center}
\includegraphics[width= 1.3 in]{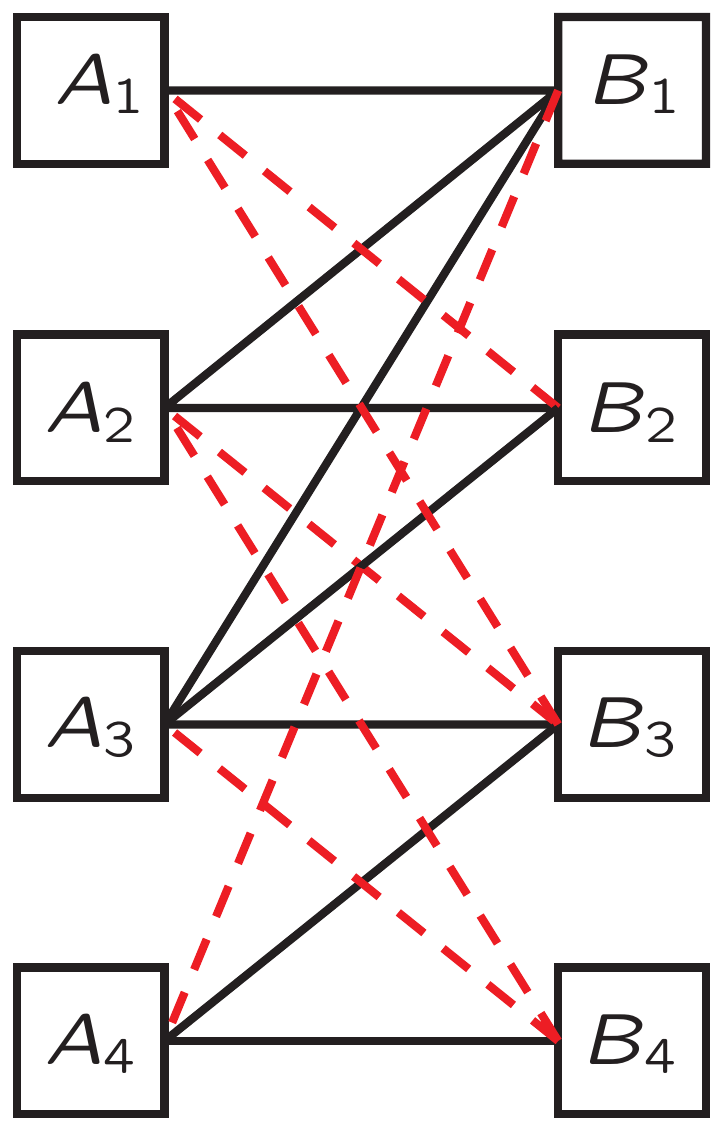}
\caption{\small A CDS instance whose linear capacity is $7/18$. The converse bound from Theorem \ref{thm:con} is not tight.}
\label{fig:ex3}
\end{center}
\end{figure}

\bigskip
The techniques from Theorem \ref{thm:con} and Theorem \ref{thm:ach} are not sufficient in general. 
Our third result is the linear capacity characterization of a CDS instance in Fig.~\ref{fig:ex3} that goes beyond previous theorems.

\begin{theorem}\label{thm:ex}
The linear capacity of the CDS instance in Fig.~\ref{fig:ex3} is $C_{\mbox{\scriptsize linear}} = 7/18$.
\end{theorem}

The proof of Theorem \ref{thm:ex} is presented in Section \ref{sec:ex}. The converse from Theorem \ref{thm:con} is not tight as $\rho = 5$ and the converse bound is $R_{\mbox{\scriptsize linear}} \leq 2/5$, which is strictly larger than $7/18$, the linear capacity. The converse proof of $R_{\mbox{\scriptsize linear}} \leq 7/18$ requires a highly non-trivial analysis of the noise and space spaces involved such that it goes well beyond the techniques from Theorem \ref{thm:con} and does not appear to admit a simple explanation (so we are not yet able to generalize it further). Once the converse bound is found, the achievable scheme follows by its guidance and falls in the general linear feasibility framework presented in Section \ref{sec:linear}.

\section{Proof of Theorem \ref{thm:con}}\label{sec:con}
The proof of Theorem \ref{thm:con} follows similarly from that of the CDS instance in Fig.~\ref{fig:ex1} considered in the previous section. We first simplify a notation that will be frequently used. For nodes $v_1, \cdots, v_i$, denote the dimension of the overlap of their noise spaces as $\alpha_{v_1\cdots v_i}$, i.e.,
\begin{eqnarray}
\alpha_{v_1\cdots v_i} \triangleq \dim( \mbox{rowspan}({\bf H}_{v_1}) \cap \cdots \cap \mbox{rowspan}({\bf H}_{v_i}) ). \label{eq:a12}
\end{eqnarray}

Consider any CDS instance $G_f(V,E)$, where $\rho \neq +\infty$ and focus on an internal qualified edge $e$ in an unqualified path $P$ such that $\rho(e, P) = \rho$. Then the connected edge cover $M$ for nodes $V_P$ in $P$ contains $\rho$ edges and $\rho+1$ nodes, denoted as $V_M = \{v_1, v_2, \cdots, v_{\rho+1}\} \subset V$. Note that such $e, P, M$ are guaranteed to exist as $\rho \neq +\infty$ and according to the definition of $\rho$, the connected edge cover $M$ attains the minimal cardinality so that $M$ is a spanning tree of the nodes $V_M$.

Start with the internal qualified edge $e$ in $M$, say $e = \{v_{i_1}, v_{i_2}\} \subset M, i_1, i_2 \in \{1,2, \cdots, \rho+1\}$. 
As $M$ is connected, there must exist a node $v_{i_3} \in V_M, i_3 \notin \{i_1, i_2\}$ and a node $u_1 \in \{v_{i_1}, v_{i_2}\}$ such that $\{u_1, v_{i_3}\}$ is a qualified edge. Then from sub-modularity, we have
\begin{eqnarray}
\alpha_{v_{i_1}v_{i_2}v_{i_3}} 
\geq \alpha_{v_{i_1}v_{i_2}} + \alpha_{u_1v_{i_3}} - N. \label{eq:a123}
\end{eqnarray}

Then we proceed similarly to find $v_{i_4} \in V_M, i_4 \notin \{i_1, i_2, i_3\}$ such that $\{u_2, v_{i_4}\}$ is a qualified edge, where $u_2 \in \{v_{i_1}, v_{i_2}, v_{i_3}\}$. Again from sub-modularity, we have
\begin{eqnarray}
\alpha_{v_{i_1}v_{i_2}v_{i_3}v_{i_4}} &\geq& \alpha_{v_{i_1}v_{i_2}v_{i_3}} + \alpha_{u_2v_{i_4}} - N \\
&\overset{(\ref{eq:a123})}{\geq}& \alpha_{v_{i_1}v_{i_2}} + \alpha_{u_1v_{i_3}} + \alpha_{u_2v_{i_4}} - 2N.
\end{eqnarray}
Continue this procedure, i.e., we include one node $v_{i_j} \in V_M, i_j \notin \{i_1, \cdots, i_{j-1}\}, j \in \{5,\cdots, \rho+1\}$ at one time such that $\{u_{j-2}, v_{i_{j}}\} \in M$ and $u_{j-2} \in \{v_{i_1}, \cdots, v_{i_{j-1}}\}$. Then we have
\begin{eqnarray}
\alpha_{v_{i_1}v_{i_2}\cdots v_{i_{\rho+1}}} &\geq& \alpha_{v_{i_1}\cdots v_{i_\rho}} + \alpha_{u_{\rho-1} v_{i_{\rho+1}}} - N \\
&\geq& \cdots \\
&\geq& \alpha_{v_{i_1}v_{i_2}} + \alpha_{u_1v_{i_3}} + \alpha_{u_2v_{i_4}} + \cdots + \alpha_{u_{\rho-1}v_{i_{\rho+1}}} - (\rho-1)N. 
\label{eq:v1v2v3vp} 
\end{eqnarray}
Note that $i_1, \cdots, i_{\rho+1}$ are distinct so that $V_M = \{v_1, \cdots, v_{\rho+1}\} = \{v_{i_1}, \cdots, v_{i_{\rho+1}}\}$.

As the $\rho+1$ noise spaces have an overlap of dimension $\alpha_{v_{i_1}v_{i_2}\cdots v_{i_{\rho+1}}}$, there exist $\rho+1$ projection matrices ${\bf P}^{\cap}_{v_{i_1}}, \cdots, {\bf P}^{\cap}_{v_{i_{\rho+1}}}$ of rank $\alpha_{v_{i_1}v_{i_2}\cdots v_{i_{\rho+1}}}$ each such that
\begin{eqnarray}
{\bf P}^{\cap}_{v_{i_1}}{\bf H}_{v_{i_1}} = {\bf P}^{\cap}_{v_{i_2}}{\bf H}_{v_{i_2}} = \cdots = {\bf P}^{\cap}_{v_{i_{\rho+1}}}{\bf H}_{v_{i_{\rho+1}}}. \label{eq:s1}
\end{eqnarray} 

Next, switch focus to the unqualified path $P$. Consider the nodes $V_P \subset V_M$ and denote $V_P =\{v_{i_1}, v_{j_1}, v_{j_2}, \cdots, v_{j_{|V_P| - 2}}, v_{i_2}\} \subset \{v_{i_1}, v_{i_2}, \cdots, v_{i_{\rho+1}}\} = V_M$ such that $\{v_{i_1}, v_{j_1}\}$, $\{v_{j_1}, v_{j_2}\}$, $\cdots$, $\{v_{j_{|V_P| - 2}}, v_{i_2}\}$ are unqualified edges. By (\ref{eq:signal}), i.e., the signal alignment constraint from Lemma \ref{lemma:align}, and (\ref{eq:s1}), we have
\begin{eqnarray}
&& {\bf P}^{\cap}_{v_{i_1}}{\bf F}_{v_{i_1}} = {\bf P}^{\cap}_{v_{j_1}}{\bf F}_{v_{j_1}} = \cdots = {\bf P}^{\cap}_{v_{j_{|V_P| - 2}}}{\bf F}_{v_{j_{|V_P| - 2}}} = {\bf P}^{\cap}_{v_{i_2}}{\bf F}_{v_{i_2}} \notag \\
&\Rightarrow&{\bf P}^{\cap}_{v_{i_1}}{\bf F}_{v_{i_1}} = {\bf P}^{\cap}_{v_{i_2}}{\bf F}_{v_{i_2}}.
\label{eq:v1v2}
\end{eqnarray}

Finally, consider the internal qualified edge $e = \{v_{i_1}, v_{i_2}\}$ and identify the noise overlap through matrices ${\bf P}_{v_{i_1}}, {\bf P}_{v_{i_2}}$ that have rank $\alpha_{v_{i_1}, v_{i_2}}$, i.e., ${\bf P}_{v_{i_1}} {\bf H}_{v_{i_1}} = {\bf P}_{v_{i_2}} {\bf H}_{v_{i_2}}$. Noting that $\mbox{rowspan}({\bf P}^\cap_{v_{i_1}})$ is a subspace of $\mbox{rowspan}({\bf P}_{v_{i_1}})$, we set
\begin{eqnarray}
{\bf P}^\cap_{v_{i_1}} = {\bf P}_{v_{i_1}}(1: \alpha_{v_{i_1}v_{i_2}\cdots v_{i_{\rho+1}}}, :), ~~{\bf P}^\cap_{v_{i_2}} = {\bf P}_{v_{i_2}}(1: \alpha_{v_{i_1}v_{i_2}\cdots v_{i_{\rho+1}}}, :) \label{eq:s2}
\end{eqnarray}
without loss of generality, i.e., the first $\alpha_{v_{i_1}v_{i_2}\cdots v_{i_{\rho+1}}}$ rows of ${\bf P}_{v_{i_1}}$ are ${\bf P}^\cap_{v_{i_1}}$.
Then from the correctness constraint (\ref{eq:dec}) for qualified edge $e = \{v_{i_1}, v_{i_2}\}$, we have
\begin{eqnarray}
L &\overset{(\ref{eq:dec})}{=}&\mbox{rank}\left({\bf P}_{v_{i_1}} {\bf F}_{v_{i_1}} - {\bf P}_{v_{i_2}} {\bf F}_{v_{i_2}}\right) \\
&\overset{(\ref{eq:v1v2}) (\ref{eq:s2})}{=}& \mbox{rank} \left( {\bf P}_{v_{i_1}} (\alpha_{v_{i_1}v_{i_2}\cdots v_{i_{\rho+1}}} +1:\alpha_{v_{1}v_{2}}, :) {\bf F}_{v_{i_1}} - {\bf P}_{v_{i_2}}(\alpha_{v_{i_1}v_{i_2}\cdots v_{i_{\rho+1}}} +1:\alpha_{v_{1}v_{2}}, :) {\bf F}_{v_{i_2}} \right) \notag\\
&&\\
&\leq& \alpha_{v_{i_1}v_{i_2}} - \alpha_{v_{i_1}v_{i_2}\cdots v_{i_{\rho+1}}} \\
&\overset{(\ref{eq:v1v2v3vp})}{\leq}&\alpha_{v_{i_1}v_{i_2}} - \left( \alpha_{v_{i_1}v_{i_2}} + \alpha_{u_1v_{i_3}} + \alpha_{u_2v_{i_4}} + \cdots + \alpha_{u_{\rho-1}v_{i_{\rho+1}}} - (\rho-1)N \right)\\
&=& (\rho-1)N - \left(\alpha_{u_1v_{i_3}} + \alpha_{u_2v_{i_4}} + \cdots + \alpha_{u_{\rho-1}v_{i_{\rho+1}}} \right) \\
&\overset{(\ref{eq:noise})}{\leq}& (\rho-1)N - (\rho-1)L \\
\Rightarrow ~ \rho L &\leq& (\rho-1)N  ~~~ \Rightarrow ~~ R_{\mbox{\scriptsize linear}} = L/(2N) ~\leq (\rho-1)/(2 \rho).
\end{eqnarray}
The proof of the linear converse bound in Theorem \ref{thm:con} is thus complete.

\section{Proof of Theorem \ref{thm:ach}}\label{sec:ach}
In this section, we present a vector linear CDS scheme that achieves rate $(\rho - 1)/(2\rho)$ as long as each qualified component of the CDS instance is either a path or a cycle. Recall that $\rho$ is the minimum connected edge cover number defined in Definition \ref{def:ec}.
Specifically, we set $L = \rho - 1$, i.e., each secret has $L$ symbols $S = (s_1, \cdots, s_{\rho-1})$ from $\mathbb{F}_p$ and $N = \rho$, i.e., each signal (node) $v$ has $N$ symbols from $\mathbb{F}_p$. We assume that $p$ is a prime number that is no smaller than $2\rho - 2$.

To prepare for the achievable scheme, we first define $L = \rho-1$ generic linear combinations $l_1, \cdots, l_{\rho-1}$ of the secret symbols.
\begin{eqnarray}
(l_1; \cdots; l_{\rho-1})_{(\rho-1) \times 1} &=& {\bf C}_{(\rho-1) \times (\rho-1)} \times (s_1; \cdots, s_{\rho-1})_{(\rho-1) \times 1} \notag\\
{\bf C}_{(\rho-1) \times (\rho-1)} (i, j) &=& \frac{1}{x_i - y_j}, i,j \in \{1,\cdots, \rho-1\} \label{eq:cc}
\end{eqnarray}
where $x_i, y_j$ are distinct elements from $\mathbb{F}_{p}$ (the existence is guaranteed by the fact that the field size $p$ is no smaller than $2\rho-2$), so ${\bf C}_{(\rho-1) \times (\rho-1)}$ is a Cauchy matrix whose every square sub-matrix has full rank \cite{Schechter}.

Consider any CDS instance $G_f(V,E)$ such that the minimum connected edge cover number for any internal qualified edge is $\rho$. Suppose the instance contains $Q$ qualified components, where each qualified component is either a path or a cycle of qualified edges. Denote the node set of the $q$-th qualified component by $V^q, q \in \{1,\cdots, Q\}$ such that $V = V^1 \cup \cdots \cup V^Q$. For each qualified component, we will use independent uniform i.i.d. noise symbols from $\mathbb{F}_p$, denoted as $z^q = (z^q_0, z^q_1, z^q_2, \cdots)$. The exact number of noise symbols used in each $z^q$ will be specified when we give the scheme and is indicated by the subscript. So $Z = (z^1, \cdots, z^Q)$.
We are now ready to specify the signal design.

\begin{enumerate}
\item Consider each qualified component sequentially. If the $q$-th qualified component is a path, go to 2; otherwise the $q$-th qualified component is a cycle, go to 3.
\item The nodes $V^q$ form a qualified path. Denote $V^q = \{v^q_1, \cdots, v^{q}_{|V^q|} \}$. Suppose $\{v^q_1, v^q_2\}$, $\{v^q_2, v^q_3\}$, $\cdots$, $\{v^q_{|V^q|-1}, v^q_{|V^q|}\}$ are qualified edges, i.e., we interpret $v^q_1$ as the first node and $v^q_{|V^q|}$ as the end node of the path. 
\begin{enumerate}
\item Assign the noise variables in a sequential manner as follows.
\begin{eqnarray}
v^q_1 = (z^q_0, z^q_1, \cdots, z^q_{\rho-1}), v^q_2 = (z^q_1, z^q_2, \cdots, z^q_{\rho}), \cdots, v^{q}_{|V^q|} = (z^q_{|V^q|-1}, \cdots, z^q_{|V^q|+\rho-2}). \label{eq:n1} 
\end{eqnarray}
\item We now describe how to the include the secret symbols to each node. Consider the nodes that contain each noise symbol $z^q_1, \cdots, z^q_{|V^q|+\rho-2}$ sequentially ($z^q_0$ will not be used) and the induced subgraph formed by these nodes. Note that each noise symbol $z^q_j, j \in \{1, \cdots, |V^q|+\rho-2\}$ appears at no more than $\rho$ nodes and denote the induced subgraph by $G^q_{j} \subset G_f$. Suppose $G_j^{q}$ contains $K^q_j$ unqualified components\footnote{A node that connects to no unqualified edge is a trivial unqualified component. As there are at most $\rho$ nodes in $G^q_{j}$, we have that $K^q_j \leq \rho$.}, each of which is considered sequentially as follows.
\begin{eqnarray}
&& \mbox{For each node $v^q_i$ in the $k$-th unqualified component of $G^q_{j}, k \in \{1,\cdots, K^q_j\},$} \notag\\
&& \mbox{$j \in \{1, \cdots, |V^q|+\rho-2\}$, replace $z_j^q$ by $k \times s_{j~\mbox{\scriptsize mod}~(\rho-1)} + z_j^q$.} \label{eq:si1}
\end{eqnarray}
Note that in $s_{j~\mbox{\scriptsize mod}~(\rho-1)}$, the subscript is defined over $\{1,\cdots, \rho-1\}$ as secret symbols are $s_1,\cdots, s_{\rho-1}$, i.e., $\{1\}, \cdots, \{\rho-1\}$ are the representative of the equivalent classes of the modulo $\rho-1$ function.
The signal assignment is complete for the path case.
\end{enumerate}
\item The nodes $V^q$ form a qualified cycle. Denote $V^q = \{v^q_1, \cdots, v^{q}_{|V^q|} \}$, and suppose $\{v^q_1, v^q_2\}$, $\{v^q_2, v^q_3\}$, $\cdots$, $\{v^q_{|V^q|-1}, v^q_{|V^q|}\}, \{v^q_{|V^q|}, v^q_1\}$ are qualified edges. 
\begin{enumerate}
\item Assign the noise variables in the following cyclic manner.
\begin{eqnarray}
&& v^q_1 = (z^q_1, z^q_2, \cdots, z^q_{\rho}), v^q_2 = (z^q_2, z^q_3, \cdots, z^q_{\rho+1}), \cdots, \notag\\
&& v^{q}_{|V^q|-1} = (z^q_{|V_q|-1}, z^q_{|V_q|}, z^q_1, \cdots, z^q_{\rho-2}), v^{q}_{|V^q|} = (z^q_{|V_q|}, z^q_1, \cdots, z^q_{\rho-1}). \label{eq:n2}
\end{eqnarray}
\item We now describe how to the include the secret symbols to each node. Consider the nodes that contain each noise symbol $z^q_1, \cdots, z^q_{|V^q|}$ sequentially and the induced subgraph formed by these nodes. Note that each noise symbol $z^q_j, j \in \{1, \cdots, |V^q|\}$ appears at $\rho$ nodes and denote the induced subgraph by $G^q_{j} \subset G_f$. Suppose $G_j^{q}$ contains $K^q_j$ unqualified components, each of which is considered sequentially as follows.
\begin{eqnarray}
&& \mbox{For each node $v^q_i$ in the $k$-th unqualified component of $G^q_{j}, k \in \{1,\cdots, K^q_j\},$} \notag\\
&& \mbox{if $j \in \{1, \cdots, \rho-1\}$, replace $z_j^q$ by $k \times l_{j} + z_j^q$;} \notag\\
&& \mbox{otherwise $j \in \{\rho, \cdots, |V_q|\}$, replace $z_j^q$ by $k \times s_{j~\mbox{\scriptsize mod}~(\rho-1)} + z_j^q$.}  \label{eq:si2}
\end{eqnarray}
Note that similar as above, $j~\mbox{mod}~(\rho-1)$ is defined over $\{1,\cdots, \rho-1\}$.
The signal assignment is complete for the cycle case.
\end{enumerate}
\end{enumerate}

After describing the signal design for all nodes, we proceed to show that the scheme is correct and secure.

First, we prove that the correctness constraint (\ref{eq:dec}) is satisfied. All qualified edges belong to some qualified component, so it suffices to consider each qualified component. We have two cases. 
\begin{itemize}
\item
The first case is when the qualified component is a path. From the noise assignment (\ref{eq:n1}), we know that the two nodes $u,v$ in any qualified edge share $L = \rho-1$ noise symbols with consecutive subscripts. Further, according to the signal assignment (\ref{eq:si1}), these $L$ consecutive noise symbols are each mixed with one distinct secret symbol from the $L$ symbols in $S$. In addition, each shared secret symbol $s_i, i \in \{1,\cdots, L\}$ in $v$ and $u$ is multiplied by different coefficients $k$ (see (\ref{eq:si1})). We prove this claim by contradiction, i.e., suppose that the coefficients $k$ are the same. Then due to the signal assignment (\ref{eq:si1}), $e = \{u,v\}$ must be an internal qualified edge in an unqualified path $P$, and we can find a connected edge cover $M$ for the nodes in $P$ and all nodes in $M$ share one same noise symbol.
Recall from Definition \ref{def:ec} that $M$ contains $\rho(e, P)$ edges and $\rho(e,P) \geq \rho$. As a result, $M$ contains at least $\rho + 1$ nodes and these nodes share one same noise symbol, which is not possible because from the noise assignment (\ref{eq:n1}), each noise symbol only appears at $\rho$ nodes at most.
Thus the coefficients for the $L$ secret symbols in $v,u$ are all distinct and from $\{v,u\}$ we can recover $S$ with no error.
\item 
The second case is when the qualified component is a cycle, whose proof is similar to the path case.
Similarly from the noise assignment (\ref{eq:n2}), any two nodes $u,v$ in a qualified edge share $L = \rho-1$ noise symbols with cyclicly consecutive subscripts. Further, according to the signal assignment (\ref{eq:si2}), these $L$ noise symbols are each mixed with either one distinct secret symbol $s_i$ from the $L$ symbols in $S$ or one generic linear combination $l_j$. With a similar reasoning as above (due to the definition of $\rho$ and each noise appears at $\rho$ nodes), the multiplicative coefficients $k$ for $s_i, l_j$ are distinct. As $l_j$ are from a Cauchy matrix (see (\ref{eq:cc})), whose every square sub-matrix has full rank, we conclude that from $\{v,u\}$ we can obtain $L$ equations of form $s_i, l_j$ thus recover $S$ with no error.
\end{itemize}

Second, we prove that the security constraint (\ref{eq:sec}) is satisfied. We have two cases for an unqualified edge.
\begin{itemize}
\item The first case is when the two nodes $u,v$ of the unqualified edge are from the same qualified component. Security is guaranteed because in the signal assignment (\ref{eq:si1}), (\ref{eq:si2}), when the noise space overlaps, the same signal equation is assigned, i.e., signal alignment is ensured and (\ref{eq:sec}) holds.
\item The second case is when the two nodes $u,v$ of the unqualified edge are from two different qualified components. As the noise symbols $z^q, z^{q'}$ are independent for distinct qualified components, the noise spaces of $u,v$ have no overlap and (\ref{eq:sec}) trivially holds.
\end{itemize}

The proof of Theorem \ref{thm:ach} is now complete.

\section{Proof of Theorem \ref{thm:ex}}\label{sec:ex}
\begin{figure}[h]
\begin{center}
\includegraphics[width= 6.5 in]{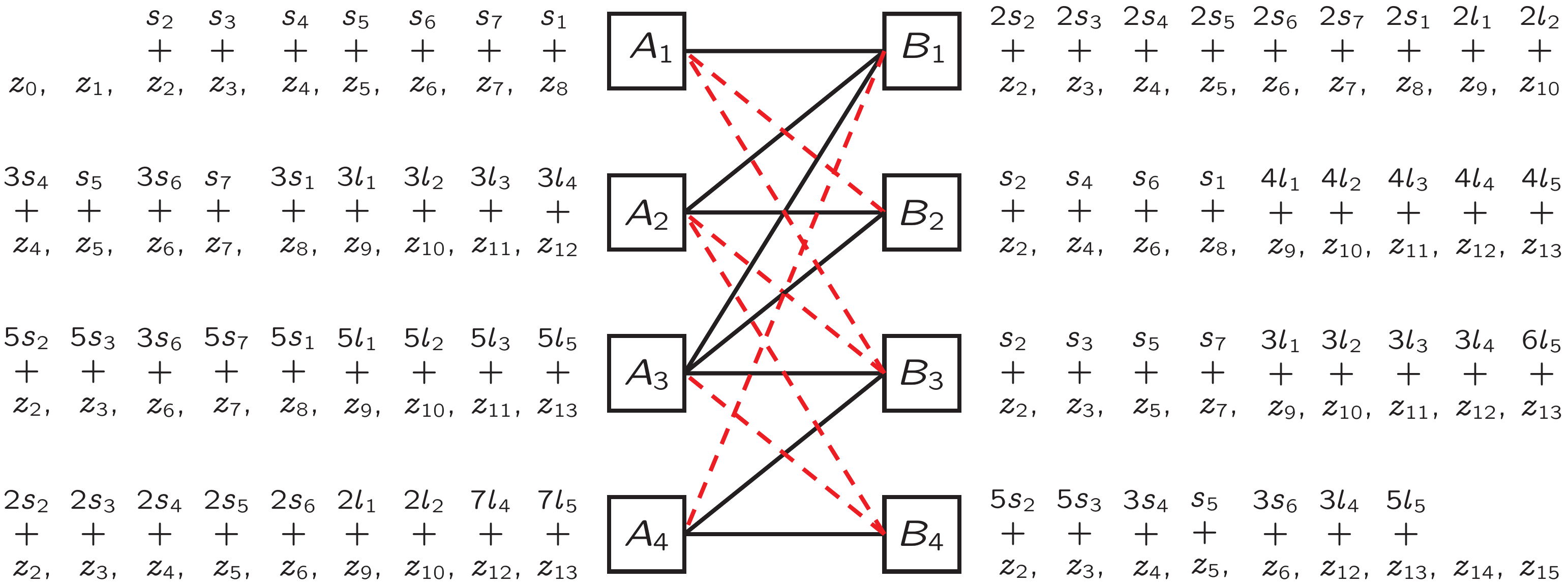}
\caption{\small A linear capacity achieving scheme with rate $7/18$. $l_i$ is a generic linear combination of $s_1,\cdots, s_7$.}
\label{fig:ex31}
\end{center}
\end{figure}

We show that the linear capacity of the CDS instance in Fig.~\ref{fig:ex3} is $7/18$. The achievable scheme is given in Fig.~\ref{fig:ex31}. The secret symbols $s_1, \cdots, s_7$ are from $\mathbb{F}_{13}$ and $l_1, \cdots, l_5$ are defined as follows.
\begin{eqnarray}
(l_1; \cdots; l_5)_{5\times 1} &=& {\bf C}_{5\times 7} \times (s_1; \cdots, s_7)_{7 \times 1} \notag \\
{\bf C}_{5\times 7}(i, j) &=& \frac{1}{x_i - y_j}, i \in \{1,\cdots,5\}, j \in \{1,\cdots,7\} 
\end{eqnarray}
where $x_i, y_j$ are distinct elements from $\mathbb{F}_{13}$ so that ${\bf C}_{5\times 7}$ is a Cauchy matrix whose every square sub-matrix has full rank.
The correctness and security constraints (\ref{eq:dec}) (\ref{eq:sec}) are straightforward to verify.

Next we provide the converse proof. Recall that $\alpha_{v_{1} \cdots v_{J}}$ denotes the dimension of the overlap of the row span of the noise precoding matrices of $v_1, \cdots, v_{J}$ and for each node, the rank of the noise precoding matrix is $N$ (see Footnote 2). 

We first give  
{an upper bound for $\alpha_{A_2, A_3}$}, where the proof is similar to that of Theorem \ref{thm:con}.
Consider nodes $A_2, A_3, B_3, A_4, B_4$. To identify their noise overlap, we may find $5$ matrices ${\bf P}_{A_2}^\cap$, ${\bf P}_{A_3}^\cap$, ${\bf P}_{B_3}^\cap$, ${\bf P}_{A_4}^\cap$, ${\bf P}_{B_4}^\cap$ of rank $\alpha_{A_2A_3B_3A_4B_4}$ each so that
\begin{eqnarray}
{\bf P}_{A_2}^\cap {\bf H}_{A_2} = {\bf P}_{A_3}^\cap {\bf H}_{A_3} = {\bf P}_{B_3}^\cap {\bf H}_{B_3} = {\bf P}_{A_4}^\cap {\bf H}_{A_4} ={\bf P}_{B_4}^\cap {\bf H}_{B_4}. \label{eq:ss1}
\end{eqnarray}

Further, $\{ \{A_3, B_4\}, \{B_4, A_2\}, \{A_2, B_3\} \}$ is an unqualified path. From (\ref{eq:signal}) and (\ref{eq:ss1}), we have
\begin{eqnarray}
{\bf P}_{A_3}^\cap {\bf F}_{A_3} = {\bf P}_{B_4}^\cap {\bf F}_{B_4} = {\bf P}_{A_2}^\cap {\bf F}_{A_2}  = {\bf P}_{B_3}^\cap {\bf F}_{B_3} ~\Rightarrow~ {\bf P}_{A_3}^\cap {\bf F}_{A_3} = {\bf P}_{B_3}^\cap {\bf F}_{B_3}. \label{eq:a3b3}
\end{eqnarray}

Consider now qualified edge $\{A_3,B_3\}$ and identify the noise overlap through ${\bf P}_{A_3}, {\bf P}_{B_3}$ of rank $\alpha_{A_3B_3}$ so that ${\bf P}_{A_3} {\bf H}_{A_3} = {\bf P}_{B_3}{\bf H}_{B_3}$. 
As $\mbox{rowspan}({\bf P}^\cap_{A_{3}})$ is a subspace of $\mbox{rowspan}({\bf P}_{A_{3}})$, without loss of generality we set
\begin{eqnarray}
{\bf P}^\cap_{A_3} = {\bf P}_{A_3}(1: \alpha_{A_2A_3B_3A_4B_4}, :), ~~{\bf P}^\cap_{B_3} = {\bf P}_{B_3}(1: \alpha_{A_2A_3B_3A_4B_4}, :).\label{eq:ss2}
\end{eqnarray}
From the correctness constraint (\ref{eq:dec}), we have
\begin{eqnarray}
L &\overset{(\ref{eq:dec})}{=}&\mbox{rank} \left({\bf P}_{A_3} {\bf F}_{A_3} - {\bf P}_{B_3}{\bf F}_{B_3} \right) \\
&\overset{(\ref{eq:a3b3}) (\ref{eq:ss2})}{=}& \mbox{rank} \left({\bf P}_{A_3}(\alpha_{A_2A_3B_3A_4B_4}+1:\alpha_{A_3B_3}, :) {\bf F}_{A_3}  - {\bf P}_{B_3}(\alpha_{A_2A_3B_3A_4B_4}+1:\alpha_{A_3B_3}, :) {\bf F}_{B_3} \right) \notag\\
&&\\
&\leq& \alpha_{A_3B_3} - \alpha_{A_2A_3B_3A_4B_4} \\
&\overset{}{\leq}&\alpha_{A_3B_3} - (\alpha_{A_2A_3} + \alpha_{A_3B_3} + \alpha_{B_3A_4} + \alpha_{A_4B_4} - 3N) \label{eq:ss} \\
&=& 3N - (\alpha_{A_2A_3} + \alpha_{B_3A_4} + \alpha_{A_4B_4}) \\
&\Rightarrow&  \alpha_{A_2A_3} \leq 3N - (\alpha_{B_3A_4} + \alpha_{A_4B_4}) - L \label{eq:ss3}
\end{eqnarray}
where (\ref{eq:ss}) follows from sub-modularity and we have obtained the desired upper bound for $\alpha_{A_2A_3}$. 

We are now ready for the final step, which is a similar chain of arguments as above. Consider $6$ nodes $A_1, B_1, A_2, B_2, A_3, B_3$ and identify their noise overlap through matrices ${\bf P}^{\cap\cap}_{A_1}$, ${\bf P}^{\cap\cap}_{B_1}$, ${\bf P}^{\cap\cap}_{A_2}$, ${\bf P}^{\cap\cap}_{B_2}$, ${\bf P}^{\cap\cap}_{A_3}$, ${\bf P}^{\cap\cap}_{B_3}$ of rank $\alpha_{A_1B_1A_2B_2A_3B_3}$ each.
\begin{eqnarray}
&& {\bf P}_{A_1}^{\cap\cap} {\bf H}_{A_1} =  {\bf P}_{B_1}^{\cap\cap} {\bf H}_{B_1} = {\bf P}_{A_2}^{\cap\cap} {\bf H}_{A_2} = {\bf P}_{B_2}^{\cap\cap} {\bf H}_{B_2} = {\bf P}_{A_3}^{\cap\cap} {\bf H}_{A_3} = {\bf P}_{B_3}^{\cap\cap} {\bf H}_{B_3} \\
&\overset{}{\Rightarrow}&  {\bf P}_{A_2}^{\cap\cap} {\bf F}_{A_2}  = {\bf P}_{B_2}^{\cap\cap} {\bf F}_{B_2} \label{eq:ss0}
\end{eqnarray}
where the last step follows from the unqualified path $\{\{A_2, B_3\}, \{B_3, A_1\}, \{A_1, B_2\}\}$ and (\ref{eq:signal}).

Consider qualified edge $\{A_2,B_2\}$ and identify the noise overlap through ${\bf P}_{A_2}, {\bf P}_{B_2}$ of rank $\alpha_{A_2B_2}$ so that ${\bf P}_{A_2} {\bf H}_{A_2} = {\bf P}_{B_2}{\bf H}_{B_2}$. Then we have 
\begin{eqnarray}
&& \mbox{rowspan}({\bf P}^{\cap\cap}_{A_{2}}) \subset \mbox{rowspan}({\bf P}_{A_{2}}), ~~\mbox{rowspan}({\bf P}^{\cap\cap}_{B_{2}}) \subset \mbox{rowspan}({\bf P}_{B_{2}})  \label{eq:ss4} \\
\Rightarrow L &\overset{(\ref{eq:dec})}{=}&\mbox{rank}({\bf P}_{A_2} {\bf F}_{A_2} - {\bf P}_{B_2}{\bf F}_{B_2}) \\
&\overset{(\ref{eq:ss0})(\ref{eq:ss4})}{\leq}& \alpha_{A_2B_2} - \alpha_{A_1B_1A_2B_2A_3B_3} \\
&\leq& \alpha_{A_2B_2} - (\alpha_{A_1B_1A_2B_2A_3} + \alpha_{A_3B_3} - N) \label{eq:ss15}  \\
&\overset{}{\leq}& \alpha_{A_2B_2} - (\alpha_{A_1B_1A_2A_3} + \alpha_{A_2B_2A_3} - \alpha_{A_2A_3} + \alpha_{A_3B_3} - N)  \label{eq:ss13} \\
&\overset{(\ref{eq:ss3})}{\leq}& \alpha_{A_2B_2} - \Big( \big(\alpha_{A_1B_1} + \alpha_{B_1A_2} + \alpha_{A_3B_1} - 2N\big) + \big(\alpha_{A_2B_2} + \alpha_{B_2A_3} - N\big) \notag\\
&&~- \big( 3N - (\alpha_{B_3A_4} + \alpha_{A_4B_4}) - L \big) + \alpha_{A_3B_3} - N\Big)  \label{eq:ss14} \\
&=& 7N - L - (\alpha_{A_1B_1} + \alpha_{B_1A_2} + \alpha_{A_3B_1} + \alpha_{B_2A_3} + \alpha_{B_3A_4} + \alpha_{A_4B_4} + \alpha_{A_3B_3}) \label{eq:ss12} \\
&\overset{(\ref{eq:noise})}{\leq}& 7N - 8L \label{eq:ss11} \\
&\Rightarrow&  R_{\mbox{\scriptsize linear}} = L/(2N) ~\leq~ 7/18.
\end{eqnarray}
where sub-modularity is repeatedly applied in (\ref{eq:ss15}), (\ref{eq:ss13}), (\ref{eq:ss14}); in (\ref{eq:ss12}), $\{v,u\}$ is a qualified edge in every $\alpha_{vu}$ term, so (\ref{eq:noise}) can be applied to obtain (\ref{eq:ss11}). The converse proof and thus the linear capacity proof of Theorem \ref{thm:ex} is complete.

\section{Discussion}
\begin{figure}[h]
\begin{center}
\includegraphics[width= 1.3 in]{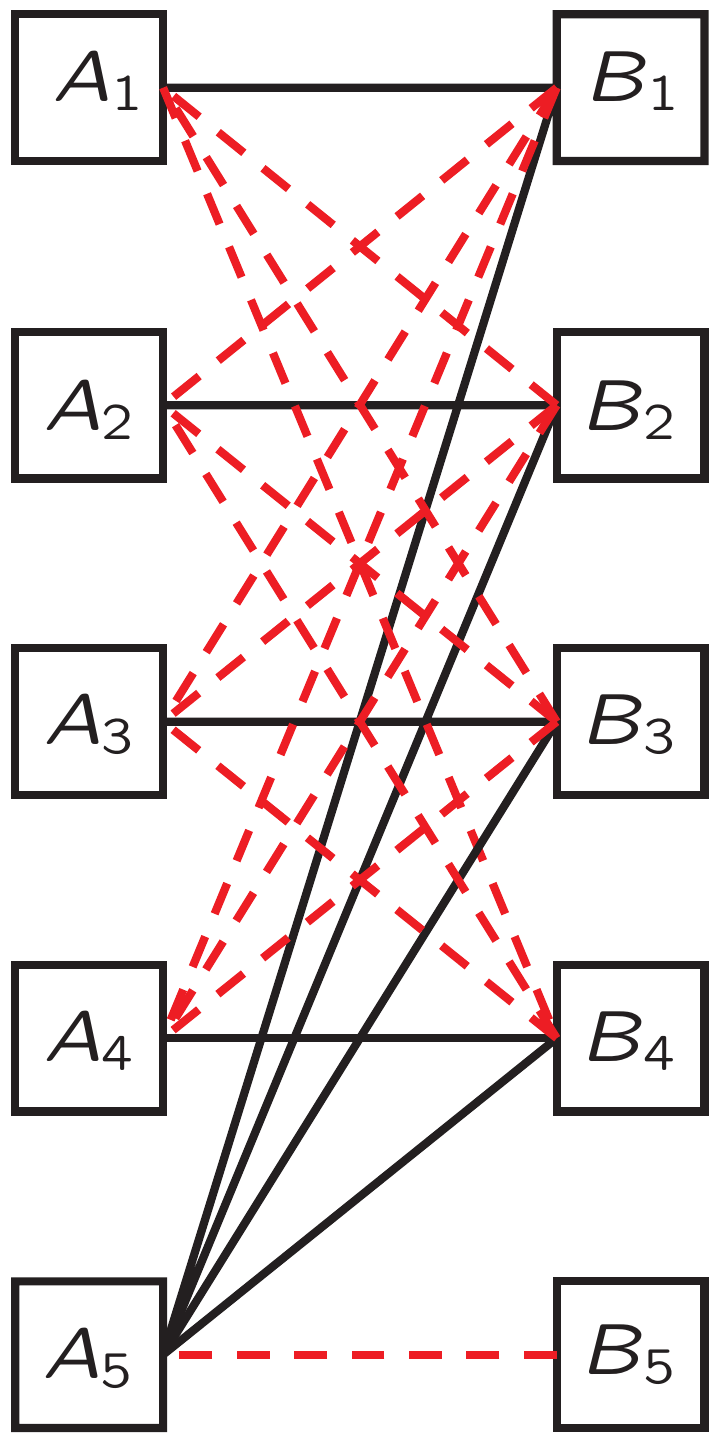}
\caption{\small A CDS instance whose linear capacity is open. The best known converse bound is from Theorem~\ref{thm:con} and is equal to $2/5$.}
\label{fig:ex4}
\end{center}
\end{figure}

In this work, we take a Shannon theoretic perspective at the canonical conditional disclosure of secrets problem to seek capacity characterizations where the secret size is allowed to approach infinity while most cryptography work focuses on the scaling of communication cost with the input size\footnote{One exception is recent work \cite{applebaum2018power}, where the amortization formulation essentially considers the same rate metric as our work. The difference is that we focus on each CDS instance and pursue exact linear capacity characterizations (so impossibility claims included) while \cite{applebaum2018power} aims at worst case rate approximation for all CDS instances.} \cite{SymPIR, Gay_Kerenidis_Wee, Applebaum_Arkis_Raykov_Vasudevan, Laur_Lipmaa, Applebaum_Vasudevan, Liu_Vaikuntanathan_Wee}. 
This Shannon theoretic perspective follows the footsteps of recent attempts in the information theory community on other cryptographic primitives \cite{Sun_Jafar_PIR, Lee_Abbe, Data_Prabhakaran_Prabhakaran, Zhou_Sun_Fu, Zhao_Sun_SMP, Sun_Anonymous, Tahmasebi_Maddah, Wang_Banawan_Ulukus, wang2020multi, Sun_CompoundSG}.
Towards this end, we further develop the noise and signal alignment approach, which is a variation of interference alignment originally studied in wireless communication networks \cite{Jafar_FnT, Jafar_TIM, Sun_Jafar_nonshannon} and is introduced in \cite{Li_Sun_CDS}, to characterize the linear capacity of a class of CDS instances, which go beyond the highest capacity scenarios found in \cite{Li_Sun_CDS}. Along the line, we identify a general linear converse bound (see Theorem \ref{thm:con}) and a linear feasibility framework that facilitates the design of linear schemes once the target rate value is fixed (see Section \ref{sec:linear}). However, these results are not sufficient to fully understand the linear capacity of CDS in general. We conclude by giving an intriguing CDS instance whose linear capacity is open (see Fig.~\ref{fig:ex4}). Note that this instance is only slightly changed from the solved instance in Fig.~\ref{fig:ex1}.

\let\url\nolinkurl
\bibliographystyle{IEEEtran}
\bibliography{Thesis}
\end{document}